\documentclass[mathpazo]{cicp}

\newcommand{\rbk}[1]{\left( #1 \right)}
\newcommand{\brc}[1]{\left\{ #1 \right\} }
\newcommand{\sbk}[1]{\left[ #1 \right]}

\newcommand{\abs}[1]{\left| #1 \right|}
\newcommand{\retn}{\nonumber \\ }

\newcommand{\dif}[2]{\frac{d #1}{d #2}}

\newcommand{\ex}[1]{e^{#1}}
\newcommand{\arrcase}[1]{\left\{ \begin{array}{ll} #1 \end{array}\right.}
\newcommand{\vc}[1]{\vec{#1}}

\begin{document}
\title{Molecular Dynamics Simulation of Plasma Surface Interaction}

\author[A.~Ito et.~al.]{Atsushi~Ito\affil{1}\comma\corrauth and Hiroaki~Nakamura\affil{2}}
\address{\affilnum{1}\ Department of Physics, Graduate School of Science, Nagoya University, Furo--cho, Chikusa--ku, Nagoya 464--8602, Japan.\\
           \affilnum{2}\ Department of Simulation Science, National Institute for Fusion Science, Oroshi--cho 322--6, Toki 509--5292, Japan.}
\emails{{\tt ito.atsushi@nifs.ac.jp} (A.~Ito), {\tt nakamura.hiroaki@nifs.ac.jp} (H.~Nakamura)}


\begin{abstract}
New interlayer intermolecular potential model was proposed and it represented ``ABAB'' staking of graphite.
Hydrogen atom sputtering on graphite surface was investigated using molecular dynamics simulation.
In the initial short time period, maintaining the flat structure of graphenes, hydrogen atoms brought about the difference interaction process in each incident energy.
The first graphene often adsorbed 5~eV hydrogen atoms and reflected almost all of 15~eV hydrogen atoms.
The hydrogen atoms which were injected at 30~eV penetrated into the inside of the graphite surface and were adsorbed between interlayer.
The desorption of $\mathrm{C}_2\mathrm{H}_2$ on the clear graphite surface was observed in only the case incident at 5~eV.
The animation of the MD simulation and radial distribution function indicated that the graphenes were peeled off one by one at regular interval. 
In common to the incident energy, the yielded molecules often had chain structures terminated by hydrogen atoms.
The erosion yield increased compared with the case of no interlayer intermolecular force.
\end{abstract}


\maketitle

\section{Introduction}

In the research into nuclear fusion, we deal with plasma surface interaction (PSI) problem \cite{Nakano,Roth,Roth2,Mech,Mech2,LHD,Garcia}.
A portion of the plasma confined in an experimental device flows into a divertor wall, which is shielded by graphite or carbon fiber composite tiles.
The incident hydrogen plasma which has weak incident energy erodes these carbon tiles in a process called chemical sputtering.
The erosion produces hydrocarbon molecules, such as $\textrm{CH}_{\mathrm{x}}$ and $\textrm{C}_2\textrm{H}_{\mathrm{x}}$, which affect the plasma confinement.
The PSI has been researched using molecular dynamics simulation (MD) \cite{Salonen,Salonen2,Alman,Marian}.

The authors investigated the PSI of graphite surfaces using the modified Brenner reactive empirical bond order (REBO) potential \cite{Nakamura}.
This MD simulation showed that if incident energy was 5 eV, many incident hydrogen atoms were absorbed by the graphite surface, while if the incident energy was 15 eV, most incident hydrogen atoms were reflected.
These behavior appear also in the case of deuterium and tritium incidence \cite{Ito_mghdt}.
This absorption and reflection can be explained by the chemical reaction between a single hydrogen atom and a single graphene \cite{Ito_ICNSP, Ito_gh1, Nakamura_PET}.
However, the number of absorbed hydrogen atoms seems to be independent of this graphite erosion because although the hydrogen atoms are absorbed by the first graphene on the surface only, multiple graphenes are destroyed simultaneously.
Because the incident hydrogen atoms pushed the graphite surface, the chemical bond between the first and second graphenes occurs.
This was the trigger of the graphite erosion.

However, the our previous work for PSI did not represent the interlayer intermolecular interaction of the graphite because a suitable model was not known.
For example, the existing model of the interlayer intermolecular interaction does not deal with ``ABAB'' stacking of the graphite structure.
Before we look into the effect of the interlayer intermolecular interaction, we should create new model of interlayer potential.

We used MD simulation to investigate hydrogen atom sputtering on graphite surface.
In \S \ref{ss:Potential}, modified Brenner REBO potential and new model of interlayer intermolecular potential are denoted.
We describe the simulation model in \S \ref{ss:SimMethod}.
In \S \ref{ss:Result}, we present simulation results.
We discuss in \S \ref{ss:Discuss}.
This paper concludes with a \S \ref{ss:Summary}.

\section{Potential Models}\label{ss:Potential}

\subsection{Modified Brenner REBO potential model}

We describe the model of Brenner reactive empirical bond order (REBO) potential \cite{Brenner} and our modification points.
This potential model is created based on Morse potential \cite{Morse},  Abell potential \cite{Abell} and Tersoff potential \cite{Tersoff1,Tersoff2}.

The potential function $U$ is defined by
\begin{eqnarray}
	U \equiv \sum_{i,j>i} \Bigg[V_{[ij]}^\mathrm{R}( r_{ij} ) - \bar{b}_{ij}(\{r\},\{\theta^\mathrm{B}\},\{\theta^\mathrm{DH}\}) V_{[ij]}^\mathrm{A}(r_{ij}) \Bigg],
	\label{eq:p7}
\end{eqnarray}
where $r_{ij}$ is the distance between the $i$--th and the $j$--th atoms.
The bond angle $\theta_{jik}^\mathrm{B}$ is the angle between the line segment which starts at the $i$--th atom and ends at the $j$--th atom  and the line segment which starts at the $i$--th atom and ends at the $k$--th atom,  as follows:
\begin{eqnarray}
	\cos\theta_{jik}^\mathrm{B} = \frac{\vc{x}_{ji} \cdot \vc{x}_{ki}}{r_{ji} r_{ki}},
	\label{eq:p4}
\end{eqnarray}
where $\vc{x}_{ij}\equiv{\vc{x}_i - \vc{x}_j}$ is the relative vector of position coordinate from the $j$--th atom to the $i$--th atom, and $r_{ij}$ is the distance between the $i$--th and the $j$--th atoms.
The dihedral angle $\theta_{kijl}^\mathrm{DH}$ is the angle between the triangle formed by the $j$--th, the $i$--th and the $k$--th atoms and the triangle formed by the $i$--th, the $j$--th and the $l$--th atoms.
The cosine function of $\theta_{kijl}^\mathrm{DH}$ is given by 
\begin{eqnarray}
	\cos\theta_{kijl}^\mathrm{DH} = \frac{\vc{x}_{ik} \times \vc{x}_{ji}}{r_{ik}r_{ji}}
		\cdot \frac{\vc{x}_{ji} \times \vc{x}_{lj}}{r_{ji}r_{lj}}.
\end{eqnarray}

The repulsive function $V_{[ij]}^\mathrm{R}( r_{ij})$ and the attractive function $V_{[ij]}^\mathrm{A}( r_{ij})$ are defined by
\begin{eqnarray}
	V_{[ij]}^\mathrm{R}( r_{ij}) &\equiv& f_{[ij]}^\mathrm{c}(r_{ij}) \rbk{1+\frac{Q_{[ij]}}{r_{ij}}}
		 A_{[ij]} \exp\rbk{-\alpha_{[ij]} r_{ij}}, \\
	V_{[ij]}^\mathrm{A}( r_{ij}) &\equiv& f_{[ij]}^\mathrm{c}(r_{ij}) \sum_{n=1}^3 B_{n[ij]} \exp\rbk{-\beta_{n[ij]} r_{ij}}.
\end{eqnarray}
The square bracket such as $[ij]$ means that each function or each parameter depends only on the species of the $i$--th and the $j$--th atoms,  for example $V_\mathrm{CC}^\mathrm{R}$, $V_\mathrm{HH}^\mathrm{R}$ and $V_\mathrm{CH}^\mathrm{R}$~($= V_\mathrm{HC}^\mathrm{R}$).
The coefficients $Q_{[ij]}$, $A_{[ij]}$, $\alpha_{[ij]}$, $B_{n[ij]}$~ and $\beta_{n[ij]}$~  are given by Table \ref{tb:tb2body}.

The cutoff function $f_{[ij]}^\mathrm{c}(r_{ij})$ determines effective ranges of the covalent bond between the $i$--th and the $j$--th atoms.
Two atoms are bound with the covalent bond if the distance $r_{ij}$ is shorter than $D_{[ij]}^\mathrm{min}$.
Two atoms are not bound with the covalent bond if the distance $r_{ij}$ is longer than $D_{[ij]}^\mathrm{max}$.
The cutoff function $f_{[ij]}^\mathrm{c}(r_{ij})$ connects the above two states smoothly as
\begin{eqnarray}
	f_{[ij]}^\mathrm{c}(x) \equiv \left\{ \begin{array}{ll} 
				1 & (x \leq D_{[ij]}^\mathrm{min}), \\
				\frac{1}{2}\sbk{1+\cos(\pi \frac{x - D_{[ij]}^\mathrm{min}}{D_{[ij]}^\mathrm{max} - D_{[ij]}^\mathrm{min}})} 
				& (D_{[ij]}^\mathrm{min} < x \leq D_{[ij]}^\mathrm{max}), \\
				0 & (x > D_{[ij]}^\mathrm{max}). \\ 
		\end{array}\right.
\end{eqnarray}
The constants $D_{[ij]}^\mathrm{min}$ and $D_{[ij]}^\mathrm{max}$ depend on the species of the two atoms (Table \ref{tb:tbDminmax}).
The cutoff function $f_{[ij]}^\mathrm{c}(r_{ij})$ distinguishes the presence of the covalent bond  between the $i$--th and the $j$--th atoms.

The potentials $V_{[ij]}^\mathrm{R}$ and $V_{[ij]}^\mathrm{A}$ in Eq. (\ref{eq:p7}) generate two--body force, because both are the function of the only distance $r_{ij}$.
The multi--body force is used instead of the effect of an electron orbital.
In this model, $\bar{b}_{ij}(\{r\},\{\theta^\mathrm{B}\},\{\theta^\mathrm{DH}\})$ in Eq. (\ref{eq:p7})   gives multi--body force and is defined by 
\begin{eqnarray}
	\bar{b}_{ij}(\{r\},\{\theta^\mathrm{B}\},\{\theta^\mathrm{DH}\})
	  &\equiv& \frac{1}{2} \Big[ b_{ij}^{\sigma-\pi}(\{r\},\{\theta^\mathrm{B}\}) + b_{ji}^{\sigma-\pi}(\{r\},\{\theta^\mathrm{B}\}) \Big] \retn
		 &&+ \Pi_{ij}^\mathrm{RC}(\{r\}) + b_{ij}^\mathrm{DH}(\{r\},\{\theta^\mathrm{DH}\}).
	\label{eq:p2}
\end{eqnarray}
The first term $\frac{1}{2}\sbk{\cdots}$ generates three--body force  except the effect of $\pi$ electrons.
The second term $\Pi_{ij}^\mathrm{RC}$ in Eq. (\ref{eq:p2}) represents the influence of radical energetics and $\pi$ bond conjugation \cite{Brenner}.
The third term $b_{ij}^\mathrm{DH}(\{r\},\{\theta^\mathrm{DH}\})$ in Eq.  (\ref{eq:p2})   derives four--body force in terms of dihedral angles.
These functions are composed of the production of cutoff functions $f_{[ij]}^\mathrm{c}(r_{ij})$.
Five-- or more--body force are generated during chemical reaction.

The function $b_{ij}^{\sigma-\pi}(\{r\},\{\theta^\mathrm{B}\})$ in Eq.  (\ref{eq:p2}) is defined by
\begin{eqnarray}
	b_{ij}^{\sigma-\pi}(\{r\},\{\theta^\mathrm{B}\})
	\equiv \Big[ 1 + \sum_{k\neq i,j} f_{[ij]}^\mathrm{c} (r_{ij})
		 \tilde{G}_i(\cos\theta_{jik}^\mathrm{B}) e^{\lambda_{[ijk]}}
	  + P_{[ij]}(N_{ij}^\mathrm{H},N_{ij}^\mathrm{C})\Big]^{-\frac{1}{2}}. \label{eq:p1}
\end{eqnarray}
The function $\tilde{G}_i$ in Eq. (\ref{eq:p1}) depends on the species of the $i$--th atom.
If $\cos\theta_{jik}^\mathrm{B} > \cos(109.47^{\circ})$ and the $i$--th atom is carbon, $\tilde{G}_i$ is defined by
\begin{eqnarray}
	\tilde{G}_i(\cos\theta_{jik}^\mathrm{B}) \equiv \sbk{1-Q_\mathrm{c}(M_i^\mathrm{t})}G_\mathrm{C}(\cos\theta_{jik}^\mathrm{B})
			 + Q_\mathrm{c}(M_i^\mathrm{t})\gamma_\mathrm{C}(\cos\theta_{jik}^\mathrm{B}).
		\label{eq:p8}
\end{eqnarray}
If $\cos\theta_{jik}^\mathrm{B} \leq \cos(109.47^{\circ})$ and the $i$--th atom is carbon, $\tilde{G}_i$ is defined by
\begin{eqnarray}
	\tilde{G}_i(\cos\theta_{jik}^\mathrm{B}) \equiv G_\mathrm{C}(\cos\theta_{jik}^\mathrm{B}).
\end{eqnarray}
And, if the $i$--th atom is hydrogen, $\tilde{G}_i$ is defined by
\begin{eqnarray}
		\tilde{G}_i(\cos\theta_{jik}^\mathrm{B}) \equiv G_\mathrm{H}(\cos\theta_{jik}^\mathrm{B}).
\end{eqnarray}
Here $G_\mathrm{C}$, $\gamma_\mathrm{C}$ and $G_\mathrm{H}$ are the sixth order polynomial spline functions.
Though the spline function $\tilde{G}_i$ needs seven coefficients, the only six coefficients are written in Brenner's paper \cite{Brenner}.
We determine the seven coefficients in table \ref{tb:tb6spGC}, \ref{tb:tb6spGammaC} and \ref{tb:tb6spGH}, respectively.
The function $Q_\mathrm{c}$ and the coordination number $M_i^\mathrm{t}$ in Eq. (\ref{eq:p8}) are defined by
\begin{eqnarray}
	Q_\mathrm{c}(x) \equiv \left\{ \begin{array}{ll}
		1 & \rbk{x \leq 3.2}, \\
			\frac{1}{2}\sbk{1 + \cos\rbk{2\pi\rbk{x-3.2}} }  & \rbk{3.2 < x \leq 3.7}, \\
			0  & \rbk{x > 3.7}, \\
			\end{array}\right.
\end{eqnarray}
\begin{eqnarray}
	M_i^\mathrm{t} \equiv \sum_{k \neq i} f_{[ik]}^\mathrm{c}(r_{ik}).
\end{eqnarray}

The constant $\lambda_{[ijk]}$ in Eq. (\ref{eq:p1}) is  a weight to modulate a strength of three--body force, which  depends on the species of the $i$--th, the $j$--th and the $k$--th atoms.
In comparison with Brenner's former potential \cite{Bernner90}, we set constants $\lambda_{[ijk]}$ as follows:
\begin{eqnarray}
	\lambda_\mathrm{HHH} &=& 4.0, \\
	\lambda_\mathrm{CCC} &=& \lambda_\mathrm{CCH} = \lambda_\mathrm{CHC} = \lambda_\mathrm{HCC} \nonumber\\
	 &=& \lambda_\mathrm{HHC} = \lambda_\mathrm{HCH} = \lambda_\mathrm{CHH} = 0.
\end{eqnarray}

The function $P_{[ij]}$ in Eq. (\ref{eq:p1}) is required  in the case that molecules forms solid structure.
The function $P_{[ij]}$ is the bicubic spline function whose coefficients depend on the species of the $i$--th and the $j$--th atoms (Table \ref{tb:tbPij}).
The parameters $N_{ij}^\mathrm{H}$ and $N_{ij}^\mathrm{C}$ are, respectively, the number of hydrogen atoms and the number of carbon atoms bound by the $i$--th atom as follows:
\begin{eqnarray}
N_{ij}^\mathrm{H} \equiv \sum_{k \neq i,j}^\mathrm{hydrogen} f_{[ik]}^\mathrm{c}(r_{ik}), \\
N_{ij}^\mathrm{C} \equiv \sum_{k \neq i,j}^\mathrm{carbon} f_{[ik]}^\mathrm{c}(r_{ik}).
\end{eqnarray}

The second term $\Pi_{ij}^\mathrm{RC}$ in Eq.  (\ref{eq:p2}) is defined by a tricubic spline function $F_{[ij]}$ as
\begin{eqnarray}
	\Pi_{ij}^\mathrm{RC}(\{r\}) \equiv F_{[ij]} (N_{ij}^\mathrm{t},N_{ji}^\mathrm{t},N_{ij}^\mathrm{conj}),
	\label{eq:p6}
\end{eqnarray}
where the variables are defined by
\begin{eqnarray}
	N_{ij}^\mathrm{t} \equiv \sum_{k \neq i,j} f_{[ik]}^\mathrm{c}(r_{ik}), \\
	N_{ij}^\mathrm{conj} \equiv 1 + \sum_{k(\neq i,j)}^\mathrm{carbon} f_{[ik]}^\mathrm{c}(r_{ik})C_\mathrm{N}(N_{ki}^\mathrm{t})
		  + \sum_{l(\neq j,i)}^\mathrm{carbon} f_{[jl]}^\mathrm{c}(r_{jl})C_\mathrm{N}(N_{lj}^\mathrm{t}), 
		\label{eq:p9}
\end{eqnarray}
with
\begin{eqnarray}
	C_\mathrm{N}(x) \equiv \arrcase{1 & (x \leq 2), \\
				\frac{1}{2}\sbk{1+\cos(\pi (x - 2))} & (2 < x \leq 3), \\
				0 & (x > 3). \\ } 
\end{eqnarray}
The second and the third terms of the right hand of Eq. (\ref{eq:p9}) are not squared.
We note that they are squared in Brenner's original formulation \cite{Brenner}.
By this modification, a numerical error becomes smaller than Brenner's formation.
Table \ref{tb:tbFij} shows the revised coefficients for $F_{[ij]}$.

The third term $b_{ij}^\mathrm{DH}(\{r\},\{\theta^\mathrm{DH}\})$ in Eq. (\ref{eq:p2}) is defined by
\begin{eqnarray}
	b_{ij}^\mathrm{DH}(\{r\},\{\theta^\mathrm{DH}\}) \equiv T_{[ij]}(N_{ij}^\mathrm{t},N_{ji}^\mathrm{t},N_{ij}^\mathrm{conj})
		 \sbk{\sum_{k\neq i,j} \sum_{l\neq j,i}
			\rbk{1-\cos^2\theta_{kijl}^\mathrm{DH}}f_{[ik]}^\mathrm{c}(r_{ik})f_{[jl]}^\mathrm{c}(r_{jl})},
\end{eqnarray}
where $T_{[ij]}$ is a tricubic spline function  and has the same variables as $F_{[ij]}$ in Eq. (\ref{eq:p6}).
The coefficients for $T_{[ij]}$ is also revised due to the modified $N_{ij}^\mathrm{conj}$~(Table \ref{tb:tbTij}).
In the present simulation, the function $T_{[ij]}$ becomes $T_\mathrm{CC}(2,2,5)$ for a perfect crystal graphene, and becomes $T_\mathrm{CC}(2,2,3)$ or $T_\mathrm{CC}(2,2,4)$ when a hydrogen atom is absorbed.

The time step should be smaller than that of general CMD.
To keep numerical error small, we set $5 \times 10^{-18} \mathrm{~s}$ in the present simulation because the potential model has complex form by cutoff functions and spline function.

\subsection{Interlayer intermolecular potential}

In the research for Interlayer intermolecular force, the binding energy has been well investigated.
However, experimental data are not nearly enough and ab--initio calculation cannot give us correct results yet \cite{Hasegawa}.
Especially, information of the repulsion of the interlayer is hardly reported.
Therefore, now, we have no other choice to create the potential model artificially.

\begin{figure}
	\centering
	\resizebox{\linewidth}{!}{\includegraphics{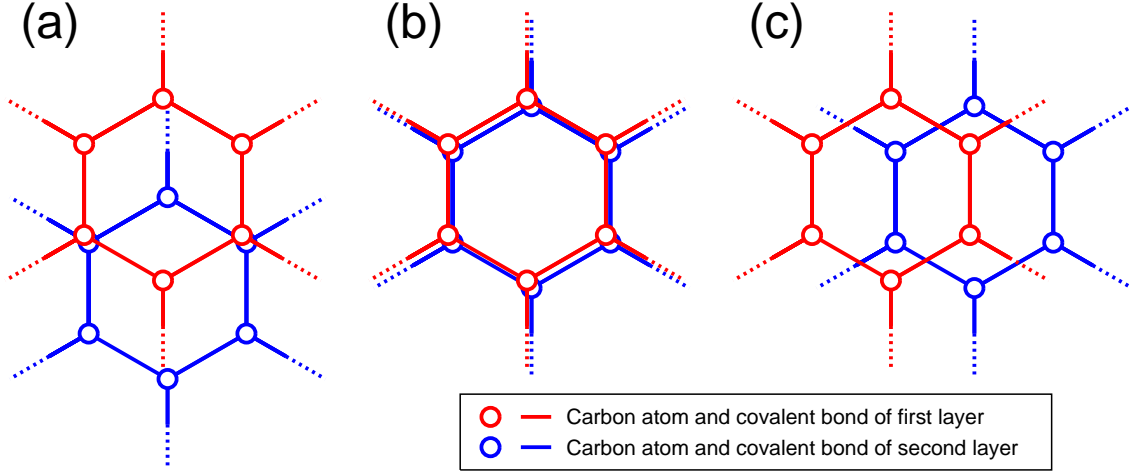}}
	\caption{The three type of stacking of the graphite}
	\label{fig:shiftlayer}
\end{figure}

\begin{figure}
	\centering
	\begin{tabular}{cc}
		\resizebox{0.5\linewidth}{!}{\includegraphics{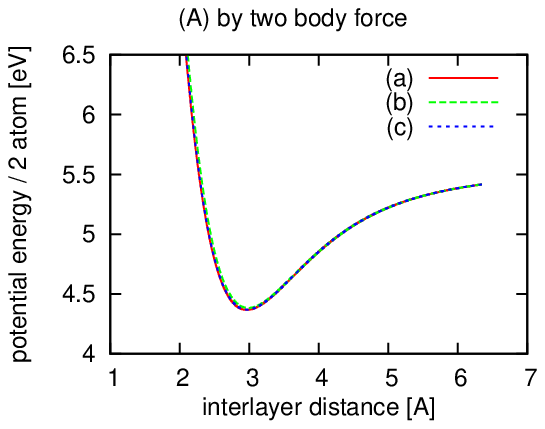}} &
		\resizebox{0.5\linewidth}{!}{\includegraphics{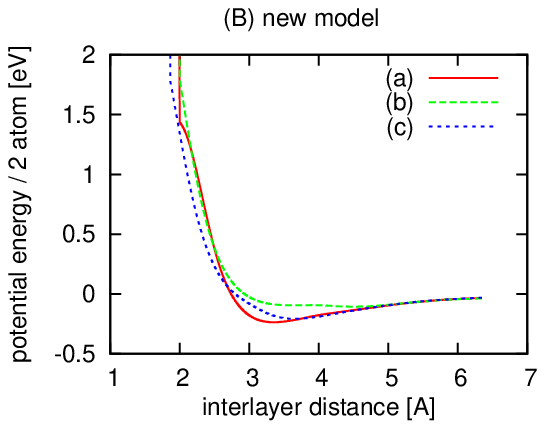}} \\
	\end{tabular}
	\caption{The interlayer potential energy by using two body force only (A) and present new model (B).
	The symbols (a), (b) and (c) correspond to the three type of stacking of the graphite in Fig. \ref{fig:shiftlayer}.}
	\label{fig:potIL}
\end{figure}

We propose new interlayer intermolecular interaction potential model for graphene layers.
First, simple intermolecular potential function between carbon atoms is defined by
\begin{eqnarray}
	V_\mathrm{IL}(r) = A \brc{\frac{n}{\alpha} \ex{-\alpha\rbk{\frac{r}{c} - 1}} - \rbk{\frac{c}{r}}^n },
	\label{eq:simpleV}
\end{eqnarray}
where $r$ is the distance between two carbon atoms, $n$ is the exponent of attraction, and $A, \alpha, c$ are the parameters to determine binding energy.
If $n > \alpha$, the potential function has a local maximum of positive energy on $r=c$.
Though we tried modelling the interlayer intermolecular potential, the challenge fell through.
Figure \ref{fig:potIL}(A) shows a potential function which consists of simple intermolecular potential of Eq. {eq:simpleV}.
Such as this potential model, we hardly produce the difference of the potential minimum energy between the three type of stacking of Fig. \ref{fig:shiftlayer}.
We consider that the difficulty comes from the use of only two body force.
The attractive interaction is regard as two body interaction historically, such as Lennard--Jones potential, Morse potential and their combination Eq. {eq:simpleV}, because it is effective in the long range.
However, the repulsive interaction is effective in the short range.
The force in the short range are provided by chemical interaction.
Therefore, the two body force is inadequate to approximate the repulsive force.

The chemical interaction is generally represented by multi--body force in the MD simulation.
Here, we propose interlayer intermolecular potential using three body force.
The product of the simple two body force $V_\mathrm{IL}(r_{ij})$ of Eq. \ref{eq:simpleV} and special cutoff function $C_{ij}$ gives us
\begin{eqnarray}
	U_\mathrm{IL} = \sum_{i,j\neq i} C_{ij} V_\mathrm{IL}(r_{ij}).
\end{eqnarray}
The special cutoff function $C_{ij}$ depends on the angles between three atoms as 
\begin{eqnarray}
	C_{ij} \equiv \frac{1}{2} &\Bigg\{ & \prod_{k \neq i} \sbk{ 1 + f_{[ij]}^\mathrm{c}(r_{ik}) \rbk{ f^\mathrm{a}(\cos \theta_{jik}) - 1 }} \\
	 &&+ \prod_{l \neq j} \sbk{ 1 + f_{[ij]}^\mathrm{c}(r_{jl})\rbk{ f^\mathrm{a}(\cos \theta_{jil}) - 1}} \Bigg\},
\end{eqnarray}
where $\vc{r}_{ij} \equiv \vc{r}_i - \vc{r}_j$, $r_{ij} = \abs{\vc{r}_{ij}}$ and $\cos \theta_{jik} = (\vc{r}_{ij} \cdot \vc{r}_{ik})/(r_{ij} r_{ik})$.
The functions $f^\mathrm{a}(\cos \theta)$ are given by
\begin{eqnarray}
	f^\mathrm{a}(\cos \theta) \equiv \left\{ \begin{array}{ll}
		1 & ( \cos \theta \leq c_\mathrm{on}), \\
			\frac {\rbk{2 \cos \theta - 3 c_\mathrm{on} + c_\mathrm{off}} \rbk {\cos \theta - c_\mathrm{off}}^2} {\rbk{c_\mathrm{off} - c_\mathrm{on}}^3} &
			(c_\mathrm{on} < \cos \theta \leq c_\mathrm{off}), \\
			0 & (\cos \theta > c_\mathrm{off}), \\
			\end{array}\right. \\
\end{eqnarray}
where $c_\mathrm{on} = 0.25$ and $c_\mathrm{off} = 0.35$.
The function $f_{[ij]}^\mathrm{c}(r)$ is equal to the cutoff function of the modified Brenner REBO potential:
\begin{eqnarray}
	f_{[ij]}^\mathrm{c}(r) \equiv \left\{ \begin{array}{ll} 
				1 & (r \leq D_{[ij]}^\mathrm{min}), \\
				\frac{1}{2}\sbk{1+\cos(\pi \frac{r - D_{[ij]}^\mathrm{min}}{D_{[ij]}^\mathrm{mar} - D_{[ij]}^\mathrm{min}})} &
				  (D_{[ij]}^\mathrm{min} < r \leq D_{[ij]}^\mathrm{max}), \\
				0 & (r > D_{[ij]}^\mathrm{max}), \\ 
		\end{array}\right.
\end{eqnarray}
where the parameters are denoted in the Table \ref{tb:tbDminmax2}
Now, we set the parameters $\alpha$ and $c$ to keep the interlayer distance 3.35~\AA~as follows: $c = 1.8$~\AA~and $\alpha = 4.84$.
If $A = 0.9961498, 2.9884494$~and~$4.980749$, the interlayer binding energy par atom becomes 20 meV, 60 meV and 100 meV, respectively (See Table \ref{tb:param}).
Figure \ref{fig:potIL}(B) show that the new interlayer intermolecular potential model provide the difference of the minimum potential energy between the three types of stacking of Fig. \ref{fig:shiftlayer}.
As a result, the structure of ``ABAB'' stacking Fig. \ref{fig:shiftlayer}(a) become the most stable state.

%

\begin{table}
\caption{The parameters of intermolecular potential. The parameters $A_{20}, A_{60}$~and~$A_{100}$ correspond to the coefficient $A$ in $V_\mathrm{IL}(r)$ and determine the binding energy of the interlayer par atom to 20 meV, 60 meV and 100 meV, respectively}
\label{tb:param}
\begin{tabular}{@{\hspace{2.0em}}l@{\hspace{2.0em}}l@{\hspace{2.0em}}l@{\hspace{2.0em}}}
\hline
	$n = 6$ & $\alpha = 4.84$ & $c = 1.8$~\AA \\
	$A_{20} = 0.9961498$ eV & $A_{60} = 2.9884494$ eV & $A_{100} = 4.980749$ \\
	$c_{on} = 0.25$ & $c_{on} = 0.35$ & ~ \\
\hline
\end{tabular}
\end{table}

\begin{table}
\caption{The constants for the cutoff function $f_{[ij]}^\mathrm{c}(r_{ij})$.
 They depend on the species of the $i$--th and the $j$--th atoms.
}

\label{tb:tbDminmax2}
\begin{tabular}{@{\hspace{2.0em}}c@{\hspace{2.0em}}c@{\hspace{2.0em}}c@{\hspace{2.0em}}}
\hline
	[ij] & $D_{[ij]}^\mathrm{min}$ & $D_{[ij]}^\mathrm{max}$ \\
	\hline
	CC & 1.7~\AA & 2.0~\AA \\
	CH & 1.3~\AA & 1.8~\AA \\
	HH & 1.1~\AA & 1.7~\AA \\
\hline
\end{tabular}
\end{table}

\section{Simulation Method}\label{ss:SimMethod}

The graphite which consists of eight graphenes \cite{Boehm} and has ``ABAB'' stacking were set to the center of coordinates parallel to $x$--$y$ plane.
Each graphene consisted of 160 carbon atoms measuring 2.00 nm $\times$ 2.17 nm.
The size of the simulation box in the $x$-- and $y$--directions is equal to that of the graphenes with the periodic boundary condition.
The inter--layer distance of the graphite was initially 3.35 \AA.
The carbon atoms obeyed the Maxwell--Boltzmann distribution at 300 K, initially.
During the simulation, two carbon atoms in the 8--th graphene were fixed to block the movement of whole of the graphite.
One was the center atom of the 7--th graphene from the surface, and the other was located at the boundary of the 8--th graphene.
The graphite surface was oriented to face the positive $z$--direction.

Hundreds of hydrogen atoms were injected at regular time intervals of 0.1 ps parallel to the $z$--axis.
The $z$--coordinate of the injection point was 60 \AA.
The $x$-- and $y$--coordinates of the injection point were set at random.
The initial momentum vector (0, 0, $p_0$) was was defined by
\begin{eqnarray}
	p_0 = \sqrt{2 m E_{\mathrm{I}}}, \label{eq:inip}
\end{eqnarray}
where $E_\mathrm{I}$ is the incident energy, and $m$ is the mass of the incident hydrogen atoms.

We adopt \textit{NVE} conditions, where the number of atoms, volume, and total energy are conserved, except for the addition of incident atoms and removal of outgoing atoms.
The simulation time was developed using second order symplectic integration \cite{Suzuki}.
The chemical interaction was represented by the modified Brenner REBO potential.
The interlayer intermolecular interaction was represented by the new model in \S \ref{ss:Potential}.
The interlayer binding energy is selected to 60 meV.
To keep the computational error of total energy small, the time step was $5\times10^{-18} \mathrm{~s}$.


\section{Results}\label{ss:Result}

We performed MD simulates for the three cases in which the incident energy of all hydrogen atoms are set into 5~eV, 15~eV or 30~eV.
In this section, simulation results are described with the story of PSI process.

\begin{figure}
	\centering
	\resizebox{\linewidth}{!}{\includegraphics[bb = 0 0 956 480]{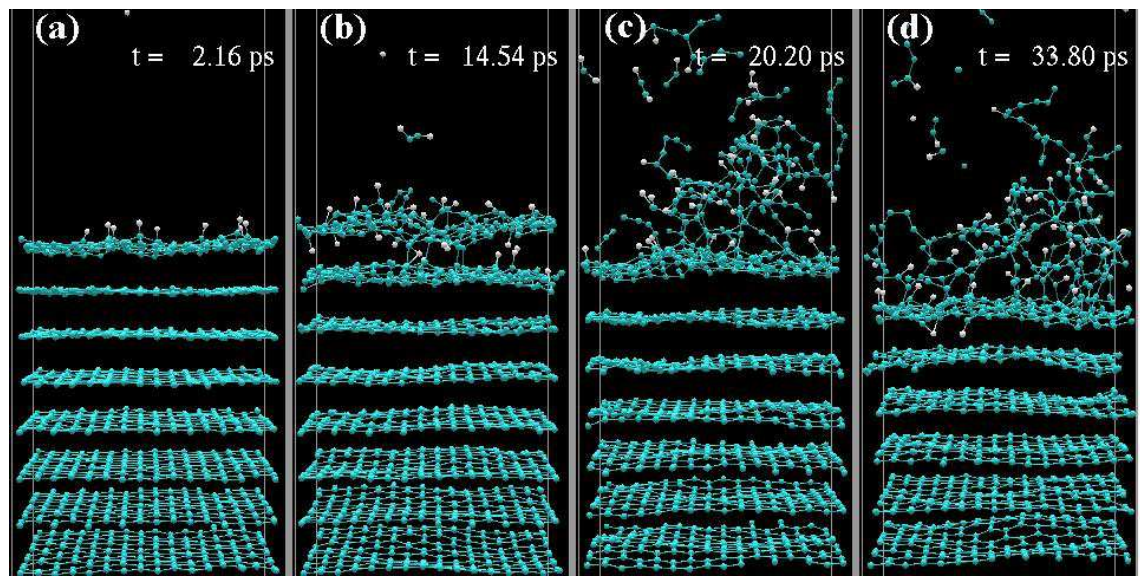}}
	\caption{The snapshot of the MD simulation for PSI in the case of the incident energy of 5~eV.
	Green and white spheres represent carbon and hydrogen atoms, respectively.}
	\label{fig:anime5eV}
\end{figure}

\begin{figure}
	\centering
	\resizebox{\linewidth}{!}{\includegraphics[bb = 0 0 956 480]{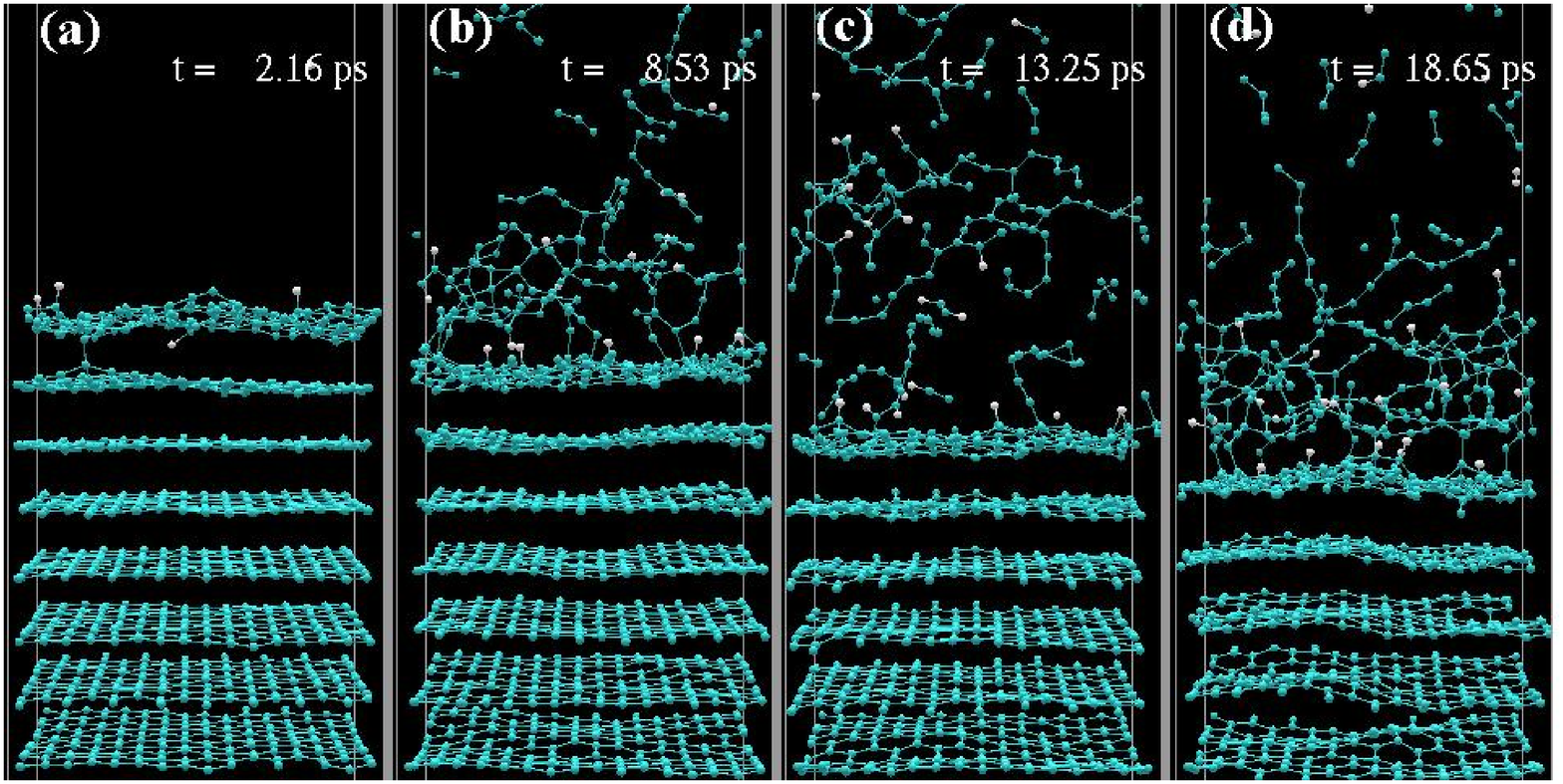}}
	\caption{The snapshot of the MD simulation for PSI in the case of the incident energy of 15~eV.
	Green and white spheres represent carbon and hydrogen atoms, respectively.}
	\label{fig:anime15eV}
\end{figure}

\begin{figure}
	\centering
	\resizebox{\linewidth}{!}{\includegraphics[bb = 0 0 956 480]{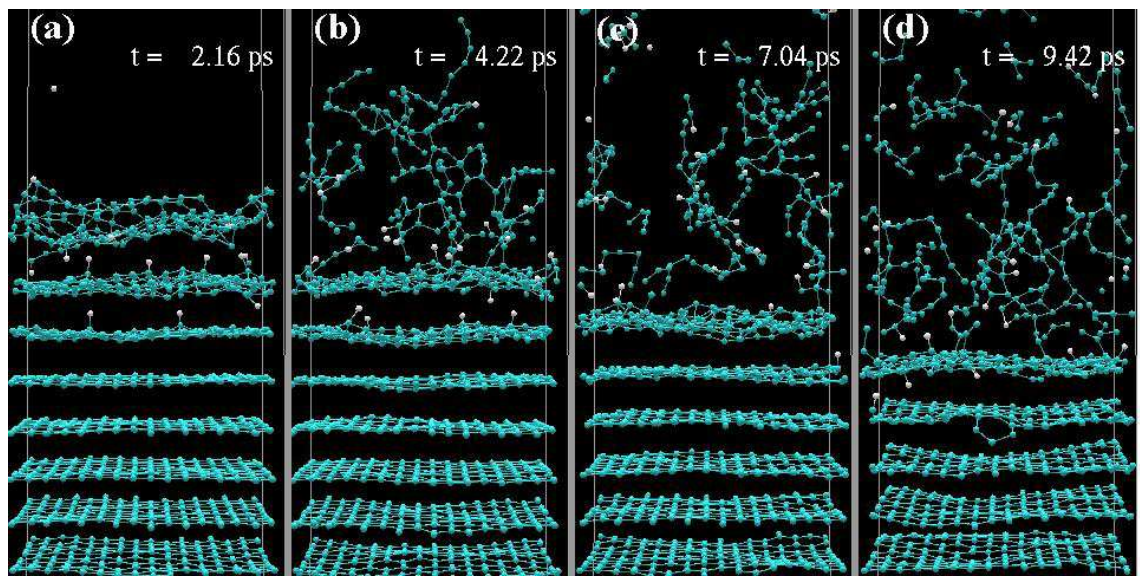}}
	\caption{The snapshot of the MD simulation for PSI in the case of the incident energy of 30~eV.
	Green and white spheres represent carbon and hydrogen atoms, respectively.}
	\label{fig:anime30eV}
\end{figure}

In the initial short time period, graphite surface were not broken.
However, the difference between the incident energy caused the difference of hydrogen atom adsorption on the graphite surface.
Figure \ref{fig:anime5eV}(a), \ref{fig:anime15eV}(a) and \ref{fig:anime30eV}(a)  show the snapshots of the MD simulation for PSI at $t=2.16$~ps, at which more than 20 hydrogen atoms had done chemical interaction with the graphite surface.
From the figures, we noticed the amount of adsorbed hydrogen atoms and adsorption sites on the graphite surface.
For incidence at 5~eV, a lot of hydrogen atoms were adsorbed by the graphite surface.
The adsorption sites are the front of the first graphene, where graphenes are numbered from surface side.
The positive and negative side of each graphene in the direction of $z$ are called front and backside, respectively.
For incidence at 15~eV, few adsorbed hydrogen atoms exist on the front of the first graphene.
The animation of the MD simulation illustrated that hydrogen atoms except for the adsorbed one were reflected by the first graphene and went back to the positive direction of $z$.
For incidence at 30~eV, a lot of hydrogen atoms were adsorbed between the first and second graphene layers, that is, the backside of the first graphene or the front of the second graphene.
A few hydrogen atom are adsorbed on the front of the third graphene.
The animation of the MD simulation of incidence at~30 eV demonstrated the following dynamics.
A lot of hydrogen atoms passed through a hexagonal opening of the first graphene, which is formed by six C--C bonds.
After that, a half of them was adsorbed on the backside of the first graphene and the others flowed between the first and second graphene layers.
When approaching the first or second graphene, the hydrogen atom was adsorbed.
The hydrogen atoms which were adsorbed by the third graphene had penetrated the first and second graphene at a stretch.
All of the hydrogen atoms which are not adsorbed by the graphite were reflected by the first graphene.
Namely, the hydrogen atoms which penetrated into the inside of the graphite surface do not go out again.
In addition, it is never seen that a hydrogen atom hits out a carbon atom and ejects it from a graphene, which is called physical sputtering.

While the graphite surface maintained the graphene sheet structure, small hydrocarbon molecules, such as $\mathrm{CH}_x$ and $\mathrm{C}_2\mathrm{H}_x$, did not occur for the incident energy of 15~eV and 30~eV.
However, for the incident energy of 5~eV, one $\mathrm{C}_2\mathrm{H}_2$ was generated keeping the first graphene flat (See Fig. \ref{fig:anime5eV}(b)).
After the atoms continued the above process, the first graphene was destroyed independent of the other graphenes.
After that, from the second sheet, the graphenes were destroyed one by one with time.
These destruction process is common to all cases of incident energy.

\begin{figure*}
	\centering
	\begin{tabular}{cc}
		\resizebox{0.5\linewidth}{!}{\includegraphics{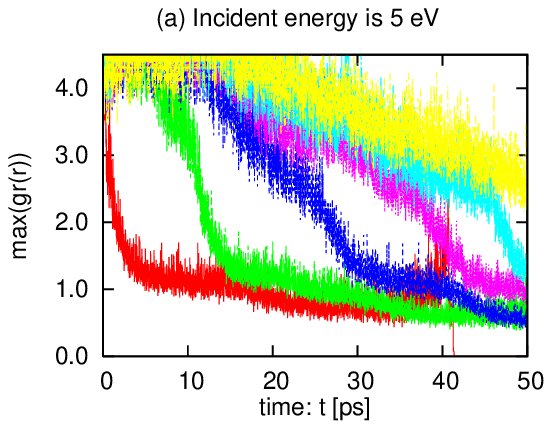}} &
		\resizebox{0.5\linewidth}{!}{\includegraphics{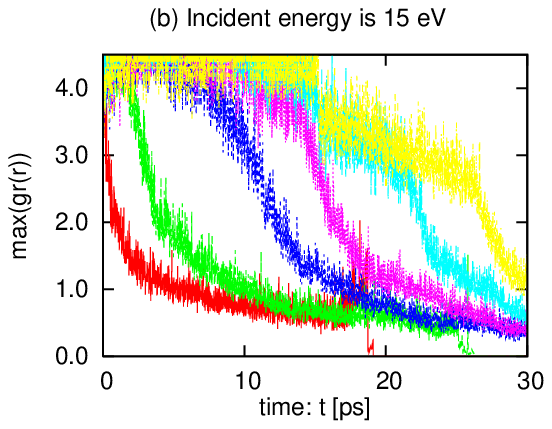}} \\
		\resizebox{0.5\linewidth}{!}{\includegraphics{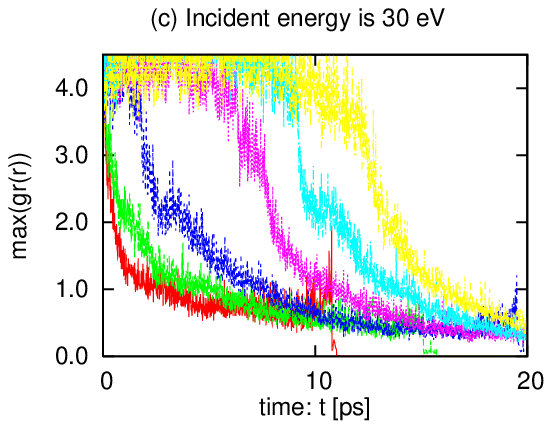}} &
		\resizebox{0.5\linewidth}{!}{\includegraphics{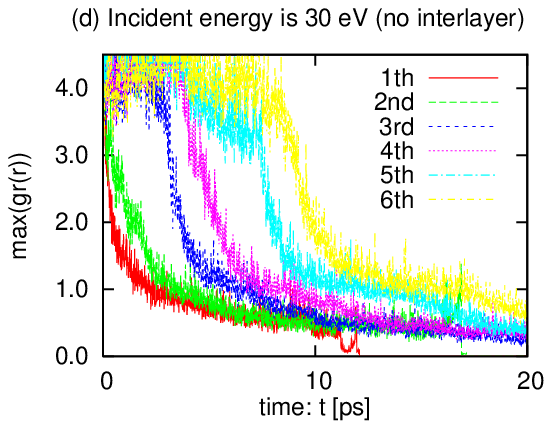}} \\
	\end{tabular}
	\caption{The maximum value of the radial distribution function $g_\mathrm{max}(r,t)$ for each graphene layers as a function of the time $t$.
	Only figure (d) is the result of the previous MD simulation which neglects the interlayer intermolecular force.}
	\label{fig:grmax}
\end{figure*}

To estimate the breakage of the graphite, radial distribution function was calculated.
However, we cannot define the three dimensional volume in this simulation model because there is no boundary in the $z$--direction.
Two dimensional radial distribution function g(r,t) of each graphene layer was defined.
First, we defined $n_i(r,t)$ as the number of the carbon atoms which are located at a distance of less than $r$ from the $i$--th carbon atom at time $t$.
The average $n(r,t)$ is then given by
\begin{eqnarray}
	n(r,t) = \sum_{i}^\mathrm{layer} \frac{n_i(r,t)}{160},
\end{eqnarray}
where $\sum_{i}^\mathrm{layer}$ means summation in only one graphene.
Consequently, the radial distribution function is given by
\begin{eqnarray}
	g(r,t) \equiv \frac{1}{4 \pi r^2} \dif{n(r,t)}{r}.
\end{eqnarray}
We calculated each graphene layer.
We plotted the maximum values of the radial distribution function $g_\mathrm{max}(r,t)$ as a function of time (see Fig. \ref{fig:grmax}).
These maximum values always demonstrated the amount of the C--C bonds of length $r = 1.42$~\AA and correspond to the number of $\textrm{sp}^2$ bonds.
The decrease of $g_\mathrm{max}(r,t)$ indicates the destruction of each graphene layer.
As the incident energy increases, the speed of the decrease of $g_\mathrm{max}(r,t)$ increases.
It seems that the fast decrease of $g_\mathrm{max}(r,t)$ of each graphene occurred at intervals.
This was roughly regular interval.

\begin{figure*}
	\centering
	\begin{tabular}{ccc}
		\resizebox{0.31\linewidth}{!}{\includegraphics{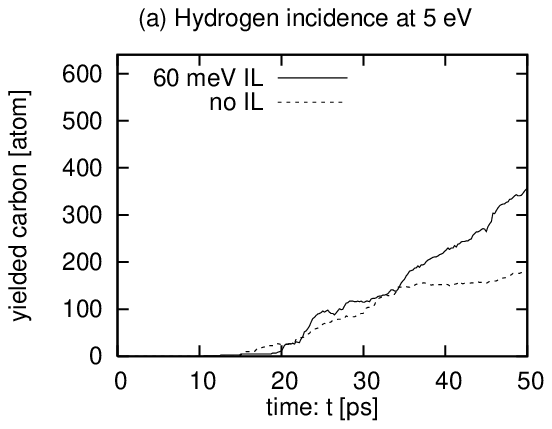}} &
		\resizebox{0.31\linewidth}{!}{\includegraphics{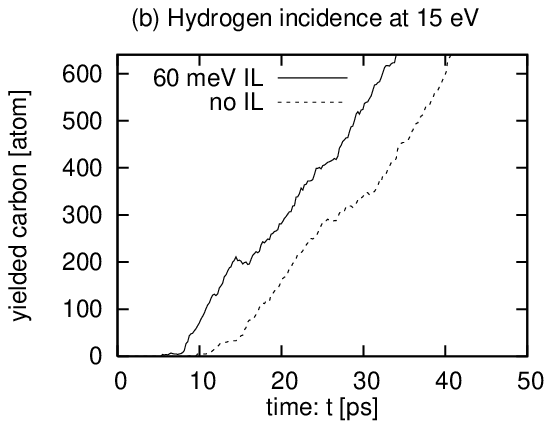}} &
		\resizebox{0.31\linewidth}{!}{\includegraphics{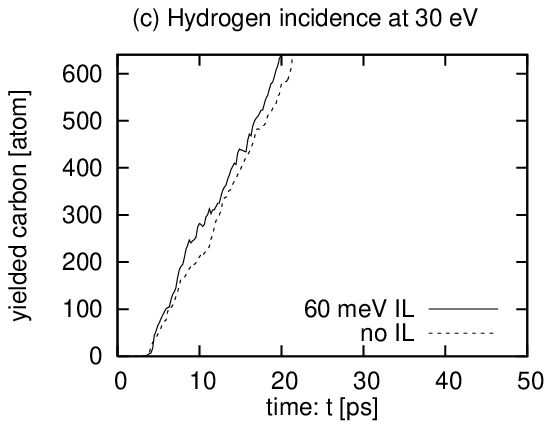}} \\
	\end{tabular}
	\caption{The erosion yields $Y$ of carbon atoms as a function of time.}
	\label{fig:yield}
\end{figure*}

Because yielded molecules repeated chemical reaction, the species of the yielded molecules was not identified, 
In common, the yielded molecules had chain structures, for example C--C--C--C--H.
Hydrogen atoms were often located at the edge of the chain molecules.
To estimate erosion yield $Y(t)$, we counted the number of the carbon atoms that moved to the region $z > 24$~\AA, where the first graphene is initially located on $z = 11.7$~\AA.
Figure \ref{fig:yield} shows the erosion yield $Y(t)$ as a function of time $t$.
The erosion yield $Y(t)$ increases with time $t$ linearly. 
As the incident energy increases, the speed of the increase of $Y(t)$ become faster and the yielded molecules started to be created earlier.
As a result of the use of the interlayer intermolecular interaction, $Y(t)$ increased.


\section{Discussion}\label{ss:Discuss}

\subsection{Initial short process}

In the initial short time period, the behavior of hydrogen atoms depended on incident energy.
This behavior can be explained by the research of the interaction between a single hydrogen atom and a single graphene because of the following three reasons.
First, the graphite surface was maintained during the initial short time period.
That is to say, after doing interaction with hydrogen atoms, the C--C bonds in the graphite are not broken and the graphenes keep their structures.
Second, because a hydrogen atom was reflected or adsorbed before the next incidence, only one hydrogen atom relates to one process.
Third, the interlayer distance of the graphite are kept about 3.35~\AA.
Because a hydrogen atom and a graphene start chemical (strong) interaction at a distance less than 1.6~\AA, in the graphite structure one hydrogen atom does not interact with two graphenes simultaneously.
We, therefore, discuss the behavior of hydrogen atoms in the initial short time period comparing with the research of the interaction between a single hydrogen atom and a single graphene.
We had already researched the interaction between a single hydrogen atom and a single graphene using MD simulation with modified Brenner REBO potential \cite{Ito_ICNSP, Ito_gh1, Nakamura_PET}.
In that simulation, a hydrogen atom was injected into a graphene vertically.
We classified interaction into three types, which is adsorption, reflection and penetration.
Moreover backside adsorption was distinguished from front adsorption.
Since the injection were repeated tens of thousands of times while changing incident position, we obtained the rates of three types interactions.
The rates of the three type interactions depend on the incident energy as follows;
If the incident energy is less than 1~eV, almost all of the interactions become the reflection due to $pi$--electron on the graphene surface.
For the incident energy from 1~eV to 7~eV, the adsorption is dominant and has a peak rate at 5 eV.
All of this adsorption is of the front of graphene surface.
For the incident energy from 7~eV to 30~eV, the reflection is dominant and has a peak at 15~eV.
As the incident energy increases from 15~eV, the rate of the penetration increases.
For the incident energy of more than 30~eV, the penetration becomes dominant.
Moreover, for the penetrate process, the hydrogen atom needs to expand a hexagonal opening of the graphene to pass through.
If the incident energy is not sufficient to expand the hexagonal hole and to leave the graphene, the hydrogen atom are adsorbed on the front or backside of the graphene.
As a result, the rate of the adsorption has a small peak around 25~eV.

In the present MD simulation, the adsorption and reflection for the incident energy of 5~eV and 15~eV are derived from the incident energy dependence of types of the interaction between a single hydrogen atom and a single graphene.
In the incident energy of 30 eV, we consider that the behavior of the hydrogen atoms is described by the combination of several type of the interactions.
The first interaction with the first graphene is similar to the interaction between a single hydrogen atom and a single graphene.
When the hydrogen atom penetrates the first graphene, it reduces its kinetic energy.
Therefore, the incident energy of next interaction with the second graphene shifts to the range for the reflection or adsorption.
Since the kinetic energy was small for the penetration, the hydrogen atom stayed in the current interlayer region.
However, because the loss of the kinetic energy due to the penetration depends on the incident point and timing, a few hydrogen atoms maintain and penetrate the second graphene.
If the hydrogen atom is driven by higher energy, it seems to go into deeper region.

In addition, the rate of interactions between a single hydrogen atom and a single graphene hardly depend on graphene temperature for the incident energy of more than 1 eV.
This fact supports the above consideration even if the hydrogen atom heat up the graphite surface. 


\subsection{Trigger of surface destruction}

We discuss the trigger of graphite surface destruction in this subsection.

In the present MD simulation, small hydrocarbons, for example $\mathrm{CH}_x$ and $\mathrm{C}_2\mathrm{H}_x$, were not created while the graphite surface maintain the graphene sheet structure.
The destruction of graphite surface seems to be melting or amorphization due to heat from the incident energy and adsorption energy.
In the case of the incident energy of 5~eV, we observed only one chemical sputtering on the clean graphite surface, which is the $\mathrm{C}_2\mathrm{H}_2$ desorption maintaining the first graphene layer flat (See Fig. \ref{fig:anime5eV}(b)).
It is considered that the incident energy flux is too high for the chemical sputtering on the clean graphite surface to occur.
However, when the interlayer interaction is used, no chemical sputtering was observed.
Therefore, we have advanced one step toward real process to have introduced the interlayer intermolecular interaction. 

Here, we had investigated a graphene erosion process in hydrogen atom gas using MD simulation \cite{Ito_gh2}.
In this previous work, the four type of C--C bonds appears.
If one carbon atom of a C--C bond is adsorbing a hydrogen atom, the C--C bond is called mono--overhang C--C bond.
If both of the two carbon atoms of a C--C bond are adsorbing a hydrogen atom on the same side of the graphene, the C--C bond is called ortho--overhang C--C bond.
If both of the two carbon atoms of a C--C bond are adsorbing a hydrogen atom on the opposite side of the graphene, the C--C bond is called para--overhang C--C bond.
If no carbon atom of a C--C bond is adsorbing a hydrogen atom, the C--C bond is called flat C--C bond.
This previous work demonstrated that the para--overhang C--C bond is the most breakable in the four type of C--C bonds.
This fact does not also support the chemical sputtering on the clean graphite surface because the para--overhang C--C bond hardly appears in the present MD simulation commonly to all case of incident energy.
Of course, our simulation cannot represent thermal effects such as thermal desorption because simulation time is so short.
However, from point of view of chemistry also, a C--C bond is not broken by the thermal effects and a carbon of a graphene cannot adsorb more than two carbon atoms.
Therefore, it is hard to create small hydrocarbon on the clean graphite surface.
We consider that the chemical sputtering occurs on the graphite surface which has many defects, edge regions or amorphous regions rather than the clean graphite surface.
The chemical sputtering on the carbon amorphous surface and the thermal desorption from a polymer were reported \cite{Salonen,Yamashiro}

Here, we note that when the interlayer intermolecular interaction is not used, the trigger of the surface destruction was the covalent bonding between the first and second graphenes.
This covalent bonding generated heat by using its binding energy and broke the graphene structure.
In the present simulation, the covalent bonds between the graphene layers hardly appear.

\subsection{Steady state of PSI}

The maximum values of the radial distribution function $g_\mathrm{max}(r,t)$ indicates the amount of $\mathrm{sp}^2$ bonds.
The decrease of $g_\mathrm{max}(r,t)$ in Fig. \ref{fig:grmax} implies that the flat structure of the graphene is broken.
As the interaction energy increases, the speed of the decrease of $g_\mathrm{max}(r,t)$ become faster and the start time when the hydrocarbon molecules are yielded became earlier.
These facts, in relation to the previous subsection, indicate that the heat of the graphite surface were derived from the incident energy regardless of chemical interaction.
The animation of the MD simulation showed that the graphenes were peeled off one by one from surface side.
In addition, it imply this fact that $g_\mathrm{max}(r,t)$ of each graphene started to decrease at regular intervals in Fig. \ref{fig:grmax}.
In the previous work in which the interlayer intermolecular force is neglected, the hydrogen atoms pressed the graphite surface because of the absence of the repulsive force between the graphene layers.
The first and second graphenes were bounded by covalent bonds.
This covalent bonding generated heat by using its binding energy and broke the graphene structure simultaneously.
From comparison, it is considered that the interlayer intermolecular interaction played a important roll of the repulsion to resist the pressure due to hydrogen atom incidence.
As a result, in the present MD simulation, the graphenes was not connected by covalent bonds and then they peeled off individually.
This is a mechanism particular to PSI because in nano graphite material science, it is thought that the attraction part of the interlayer intermolecular force is important for making a molecular structure.

The erosion yield $Y(t)$ increases with time $t$ linearly.
This linearity process is namely regard as steady state.
The steady state is also adhered by the fact that the graphenes were peeled off at regular intervals.
Of course, because the number of the graphite layers are finite, the steady state did not continue for a long time.

The present MD simulation achieved the steady state without temperature control.
Therefore, the present MD simulation perhaps differs from real PSI process.
Though some temperature control methods control exists, the problem is not solved even if the methods is used.
The temperature control methods usually brings about rapid cooling because we has to finish a cooling process in the MD simulation time, which is at most nano--seconds.
If the thermostat for temperature control acts to the graphite surface directly, the movement of the atoms are restricted.
Consequently, the chemical interaction on the graphite surface become far from a real behavior.
On the other hand, in the research of MD simulation, we often create or use potential models to achieve a realistic trajectory.
Thereby, the speed of heat transport is also realistic, that is, so slow for the time scale of the MD simulation.
As a result, if the thermostat is set to the region which is remote from the graphite surface, the heat cannot be transported to the thermostat.
We have to create a new method of the temperature control in the near future.

In the present work, because the covalent bonds between the graphene layers hardly occur, the heat of the surface is not transport to lower graphene layers.
Therefore, the erosion yield $Y(t)$ increased compared with the case in which the interlayer intermolecular force was not used.


\section{Summary}\label{ss:Summary}

The new model of the interlayer intermolecular potential to represent ``ABAB'' staking of the graphite was proposed.
We performed the MD simulation of the hydrogen atom sputtering on the graphite surface for the three cases of the incident energy of 5~eV, 15~eV or 30~eV.
In the initial short time period, keeping the graphene structure flat, the hydrogen atoms brought about the difference interaction process.
The first graphene adsorbed a lot of the hydrogen atoms which were injected at 5~eV and reflected almost all of the hydrogen atoms  which were injected at 15~eV.
The hydrogen atoms which were incident at 30~eV penetrated under the first graphene and were adsorbed between the graphene interlayer.
These process is similar to the interaction between a single hydrogen atom and a single graphene.
The $\mathrm{C}_2\mathrm{H}_2$ desorption on the clear graphite surface was observed in only the case of the incident energy of 5~eV.
However, the small hydrocarbon molecules except for the one $\mathrm{C}_2\mathrm{H}_2$ were not generated from the clear graphite surface.
We discussed this fact by using the result of the research for the graphene erosion in hydrogen atom gas.
The animation of the MD simulation and the maximum values of the radial distribution function $g_\mathrm{max}(r,t)$ for each graphene indicated that the graphenes were peeled off one by one at regular interval. 
In common to there cases of incident energy, the yielded molecules often had chain structures which is terminated by the hydrogen atoms.
The linear increase of the erosion yield $Y(t)$ is regard as the steady state of the sputtering process.
The erosion yield $Y(t)$ increased compared with the case in which the interlayer intermolecular force was not used.

\section*{Acknowledgments}
The authors acknowledge stimulating discussion with Dr. Arimichi Takayama.
The numerical simulations were carried out using the Plasma Simulator at the National Institute for Fusion Science.
This work was supported in part by a Grand--in Aid for Exploratory Research (C), 2007, No. 17540384 from the Ministry of Education, Culture, Sports, Science and Technology.
This work was also supported by National Institutes of Natural Sciences  undertaking for Forming Bases for Interdisciplinary and International Research through Cooperation Across Fields of Study, and Collaborative Research Programs (No. NIFS07KDAT012, No. NIFS07KTAT029, No. NIFS07USNN002 and No. NIFS07KEIN0091).



\begin{table*}
\caption{The parameters for the repulsive function $V_{[ij]}^\mathrm{R}$ and the attractive function $V_{[ij]}^\mathrm{A}$.
 They depend on the species of the $i$--th and the $j$--th atoms.}
\label{tb:tb2body}
\begin{tabular*}{\linewidth}{@{\extracolsep{\fill}}@{\hspace{2.0em}}cccc@{\hspace{2.0em}}}
\hline
	~ & \multicolumn{3}{c}{$[ij]$} \\
	\cline{2-4}
	Parameter & CC & HH & CH or HC \\
	\hline
	$Q_{[ij]}$ & 0.3134602960833 \AA & 0.370471487045 \AA & 0.340775728 \AA \\
	$A_{[ij]}$ & 10953.544162170 eV & 32.817355747 eV & 149.94098723 eV \\
	$\alpha_{[ij]}$ & 4.7465390606595 $\textrm{\AA}^{-1}$ & 3.536298648 $\textrm{\AA}^{-1}$ & 4.10254983 $\textrm{\AA}^{-1}$ \\
	$B_{1[ij]}$ & 12388.79197798 eV & 29.632593 eV & 32.3551866587 eV \\
	$B_{2[ij]}$ & 17.56740646509 eV & 0 eV & 0 eV \\
	$B_{3[ij]}$ & 30.71493208065 eV & 0 eV & 0 eV \\
	$\beta_{1[ij]}$ & 4.7204523127 $\textrm{\AA}^{-1}$ & 1.71589217 $\textrm{\AA}^{-1}$ & 1.43445805925 $\textrm{\AA}^{-1}$ \\
	$\beta_{2[ij]}$ & 1.4332132499 $\textrm{\AA}^{-1}$ & 0 $\textrm{\AA}^{-1}$ & 0 $\textrm{\AA}^{-1}$ \\
	$\beta_{3[ij]}$ & 1.3826912506 $\textrm{\AA}^{-1}$ & 0 $\textrm{\AA}^{-1}$ & 0 $\textrm{\AA}^{-1}$ \\
\hline
\end{tabular*}
\end{table*}

\begin{table}
\caption{The constants for the cutoff function $f_{[ij]}^\mathrm{c}(r_{ij})$.
 They depend on the species of the $i$--th and the $j$--th atoms.
}
\label{tb:tbDminmax}
\begin{tabular*}{\linewidth}{@{\extracolsep{\fill}}@{\hspace{2.0em}}ccc@{\hspace{2.0em}}}
\hline
	[ij] & $D_{[ij]}^\mathrm{min}$~(\AA) & $D_{[ij]}^\mathrm{max}$~(\AA) \\
	\hline
	CC & 1.7 & 2.0 \\
	CH & 1.3 & 1.8 \\
	HH & 1.1 & 1.7 \\
\hline
\end{tabular*}
\end{table}

\begin{table}
\caption{The parameters for the sixth order spline function $G_\mathrm{C}(\cos\theta_{jik}^\mathrm{B})$.}
\label{tb:tb6spGC}
\begin{tabular*}{\linewidth}{@{\extracolsep{\fill}}ccccc}
\hline
	$\cos\theta_{jik}^\mathrm{B}$ & $G_\mathrm{C}$ & $G_\mathrm{C}'$ & $G_\mathrm{C}''$ & $G_\mathrm{C}^{(3)}$ \\
	\hline
	$-1$ & $-$0.001 & 0.10400 & 0 & 0 \\
	$-1/2$ & 0.05280 & 0.170 & 0.370 & $-$5.232 \\ 
	$\cos(109.47^{\circ})$ & 0.09733 & 0.400 & 1.980 & 41.6140 \\ 
	1 & 8.0 & 0.23622 & $-$166.1360 & --- \\ 
\hline
\end{tabular*}
\end{table}

\begin{table}
\caption{The parameters for the sixth order spline function $\gamma_\mathrm{C}(\cos\theta_{jik}^\mathrm{B})$.}
\label{tb:tb6spGammaC}
\begin{tabular*}{\linewidth}{@{\extracolsep{\fill}}ccccc}
\hline
	$\cos\theta_{jik}^\mathrm{B}$ & $\gamma_\mathrm{C}$ & $\gamma_\mathrm{C}'$ & $\gamma_\mathrm{C}''$ & $\gamma_\mathrm{C}^{(3)}$ \\
	\hline
	$\cos(109.47^{\circ})$ & 0.09733 & 0.400 & 1.980 & $-$9.9563027\\
	1 & 1.0 & 0.78 & $-$11.3022275  & --- \\ 
\hline
\end{tabular*}
\end{table}

\begin{table}
\caption{The parameters for the sixth order spline function $G_\mathrm{H}(\cos\theta_{jik}^\mathrm{B})$.
	The parameters are determined under $\cos\theta_{jik}^\mathrm{B} = 0$.}
\label{tb:tb6spGH}
\begin{tabular*}{\linewidth}{@{\extracolsep{\fill}}@{\hspace{5.0em}}lr@{\hspace{5.0em}}}
\hline
	Parameter & Value\hspace{1.0em} \\
	\hline
	$G_\mathrm{H}(0)$       &19.06510\\
	$G_\mathrm{H}'(0)$      & 1.08822\\
	$G_\mathrm{H}''(0)$     &-1.98677\\
	$G_\mathrm{H}^{(3)}(0)$ & 8.52604\\
	$G_\mathrm{H}^{(4)}(0)$ &-6.13815\\
	$G_\mathrm{H}^{(5)}(0)$ &-5.23587\\
	$G_\mathrm{H}^{(6)}(0)$ & 4.67318\\
\hline
\end{tabular*}
\end{table}

\begin{table}
\caption{Parameters for the bicubic spline function $P_{[ij]}(N_{ij}^\mathrm{H},N_{ij}^\mathrm{C})$.
 The parameters which are not denoted are zero.
}
\label{tb:tbPij}
\begin{tabular*}{\linewidth}{@{\extracolsep{\fill}}@{\hspace{2.0em}}cc@{\hspace{2.0em}}}
\hline
	$P_{[ij]}(N_{ij}^\mathrm{H},N_{ij}^\mathrm{C})$ & \multicolumn{1}{c}{\hspace{3.0em}Value} \\ 
	\hline
$P_\mathrm{CC}(1, 1)$& 0.003026697473481 \\ 
$P_\mathrm{CC}(2, 0)$& 0.007860700254745 \\ 
$P_\mathrm{CC}(3, 0)$& 0.016125364564267 \\ 
$P_\mathrm{CC}(1, 2)$& 0.003179530830731 \\ 
$P_\mathrm{CC}(2, 1)$& 0.006326248241119 \\ 
$P_\mathrm{CH}(1, 0)$&  0.2093367328250380 \\ 
$P_\mathrm{CH}(2, 0)$& -0.064449615432525 \\ 
$P_\mathrm{CH}(3, 0)$& -0.303927546346162 \\ 
$P_\mathrm{CH}(0, 1)$&  0.01 \\ 
$P_\mathrm{CH}(0, 2)$& -0.1220421462782555 \\ 
$P_\mathrm{CH}(1, 1)$& -0.1251234006287090 \\ 
$P_\mathrm{CH}(2, 1)$& -0.298905245783 \\ 
$P_\mathrm{CH}(0, 3)$& -0.307584705066 \\ 
$P_\mathrm{CH}(1, 2)$& -0.3005291724067579 \\ 
\hline
\end{tabular*}
\end{table}

\renewcommand{\arraystretch}{0.7}
\begin{table}
\caption{Parameters for the tricubic spline function $F_{[ij]}$.
 The parameters which are not denoted are zero.
 The function $F_{[ij]}$ satisfies the following rules:
 $F_{[ij]}(N_1,N_2,N_3)=F_{[ij]}(N_2,N_1,N_3)$,
 $\partial_{N_1} F_{[ij]}(N_1,N_2,N_3)  = \partial_{N_1} F_{[ij]}(N_2,N_1,N_3)$,
 $F_{[ij]}(N_1,N_2,N_3)=F_{[ij]}(3,N_2,N_3)$~ if $N_1>3$,
 and $F_{[ij]}(N_1,N_2,N_3)=F_{[ij]}(N_1,N_2,5)$~ if $N_3>5$,
 where $\partial_{N_i} \equiv \partial / \partial N_i$.
}
\label{tb:tbFij}
\begin{tabular*}{\linewidth}{@{\extracolsep{\fill}}cccc@{\hspace{1em}}l}
\hline
	~ & \multicolumn{3}{c}{Variables} & ~ \\ 
	\cline{2-4}
	Function & $N_1$ & $N_2$ & $N_3$ & \hspace{2.0em}Value \\ 
	\hline
	$F_\mathrm{CC}(N_1,N_2,N_3)$& 1& 1& 1& \hspace{0.4525em} 0.105000 \\ 
	~ & 1& 1& 2& $-$0.0041775 \\ 
	~ & 1& 1& 3 to 5& $-$0.0160856 \\ 
	~ & 2& 2& 1& \hspace{0.4525em} 0.09444957 \\ 
	~ & 2& 2& 2& \hspace{0.4525em} 0.04632351 \\ 
	~ & 2& 2& 3& \hspace{0.4525em} 0.03088234 \\ 
	~ & 2& 2& 4& \hspace{0.4525em} 0.01544117 \\ 
	~ & 2& 2& 5& \hspace{0.4525em} 0.0 \\ 
	~ & 0& 1& 1& \hspace{0.4525em} 0.04338699 \\ 
	~ & 0& 1& 2& \hspace{0.4525em} 0.0099172158 \\ 
	~ & 0& 2& 1& \hspace{0.4525em} 0.0493976637 \\ 
	~ & 0& 2& 2& $-$0.011942669 \\ 
	~ & 0& 3& 1 to 5& $-$0.119798935 \\ 
	~ & 1& 2& 1& \hspace{0.4525em} 0.0096495698 \\ 
	~ & 1& 2& 2& \hspace{0.4525em} 0.030 \\ 
	~ & 1& 2& 3& $-$0.0200 \\ 
	~ & 1& 2& 4 to 5& $-$0.030133632 \\ 
	~ & 1& 3& 2 to 5& $-$0.124836752 \\ 
	~ & 2& 3& 1 to 5& $-$0.044709383 \\ 
	$\partial_{N_1} F_\mathrm{CC}(N_1,N_2,N_3)$ & 2& 1& 1& $-$0.052500 \\ 
	~ & 2& 1& 3 to 5& $-$0.054376 \\ 
	~ & 2& 3& 1& \hspace{0.4525em} 0.0 \\ 
	~ & 2& 3& 2 to 5& \hspace{0.4525em} 0.062418 \\ 
	$\partial_{N_3} F_\mathrm{CC}(N_1,N_2,N_3)$ & 2& 2& 4& $-$0.006618 \\ 
	~ & 1& 1& 2& $-$0.060543 \\ 
	~ & 1& 2& 3& $-$0.020044 \\ 
	$F_\mathrm{HH}(N_1,N_2,N_3)$ & 1& 1& 1& \hspace{0.4525em} 0.249831916 \\ 
	$F_\mathrm{CH}(N_1,N_2,N_3)$ & 0& 2& 3 to 5& $-$0.009047787516128811 \\ 
	~ & 1& 3& 1 to 5& $-$0.213 \\ 
	~ & 1& 2& 1 to 5& $-$0.25 \\ 
	~ & 1& 1& 1 to 5& $-$0.5 \\ 
\hline
\end{tabular*}
\end{table}

\renewcommand{\arraystretch}{1.5}
\begin{table}
\caption{Parameters for the tricubic spline function $T_\mathrm{CC}$.
 The parameters which are not denoted are zero.
 The function $T_\mathrm{CC}$ satisfies the following rule:
 $T_\mathrm{CC}(N_1,N_2,N_3)=T_\mathrm{CC}(N_1,N_2,5)$ if $N_3 > 5$.
}
\label{tb:tbTij}
\begin{tabular*}{\linewidth}{@{\extracolsep{\fill}}@{\hspace{2.0em}}ccccl@{\hspace{2.0em}}}
\hline
		~ & \multicolumn{3}{c}{Variables}  & ~ \\ 
		\cline{2-4}
		Function & $N_1$ & $N_2$ & $N_3$ & \hspace{2.0em}Value \\ 
		\hline
	$T_\mathrm{CC}(N_1, N_2, N_3)$ & 2& 2& 1& $-$0.070280085 \\ 
	~ & 2& 2& 5& $-$0.00809675 \\ 
\hline
\end{tabular*}
\end{table}

\end{document}